\documentclass[twoside,twocolumn,9pt]{article}
\usepackage{extsizes}
\usepackage[super,sort&compress,comma]{natbib} 
\usepackage[version=3]{mhchem}
\usepackage[left=1.5cm, right=1.5cm, top=1.785cm, bottom=2.0cm]{geometry}
\usepackage{balance}
\usepackage{mathptmx}
\usepackage{sectsty}
\usepackage{graphicx} 
\usepackage{lastpage}
\usepackage[format=plain,justification=justified,singlelinecheck=false,font={stretch=1.125,small,sf},labelfont=bf,labelsep=space]{caption}
\usepackage{float}
\usepackage{fancyhdr}
\usepackage{fnpos}
\usepackage[english]{babel}
\addto{\captionsenglish}{%
  
}
\usepackage{array}
\usepackage{droidsans}
\usepackage{charter}
\usepackage[T1]{fontenc}
\usepackage[usenames,dvipsnames]{xcolor}
\usepackage{setspace}
\usepackage[compact]{titlesec}
\usepackage{hyperref}
\usepackage{braket}
\usepackage{booktabs}

\usepackage{epstopdf}

\definecolor{cream}{RGB}{222,217,201}

\begin{document}

\pagestyle{fancy}
\thispagestyle{plain}
\fancypagestyle{plain}{
	\renewcommand{\headrulewidth}{0pt}
}

\makeFNbottom
\makeatletter
\renewcommand\LARGE{\@setfontsize\LARGE{15pt}{17}}
\renewcommand\Large{\@setfontsize\Large{12pt}{14}}
\renewcommand\large{\@setfontsize\large{10pt}{12}}
\renewcommand\footnotesize{\@setfontsize\footnotesize{7pt}{10}}
\makeatother

\renewcommand{\thefootnote}{\fnsymbol{footnote}}
\renewcommand\footnoterule{\vspace*{1pt}%
	\color{cream}\hrule width 3.5in height 0.4pt \color{black}\vspace*{5pt}}
\setcounter{secnumdepth}{5}

\makeatletter
\renewcommand\@biblabel[1]{#1}
\renewcommand\@makefntext[1]%
{\noindent\makebox[0pt][r]{\@thefnmark\,}#1}
\makeatother
\renewcommand{\figurename}{\small{Fig.}~}
\sectionfont{\sffamily\Large}
\subsectionfont{\normalsize}
\subsubsectionfont{\bf}
\setstretch{1.125} 
\setlength{\skip\footins}{0.8cm}
\setlength{\footnotesep}{0.25cm}
\setlength{\jot}{10pt}
\titlespacing*{\section}{0pt}{4pt}{4pt}
\titlespacing*{\subsection}{0pt}{15pt}{1pt}

\fancyfoot{}
\fancyfoot[LO,RE]{\vspace{-7.1pt}\includegraphics[height=9pt]{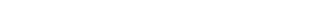}}
\fancyfoot[CO]{\vspace{-7.1pt}\hspace{13.2cm}\includegraphics{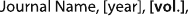}}
\fancyfoot[CE]{\vspace{-7.2pt}\hspace{-14.2cm}\includegraphics{RF}}
\fancyfoot[RO]{\footnotesize{\sffamily{1--\pageref{LastPage} ~\textbar  \hspace{2pt}\thepage}}}
\fancyfoot[LE]{\footnotesize{\sffamily{\thepage~\textbar\hspace{3.45cm} 1--\pageref{LastPage}}}}
\fancyhead{}
\renewcommand{\headrulewidth}{0pt}
\renewcommand{\footrulewidth}{0pt}
\setlength{\arrayrulewidth}{1pt}
\setlength{\columnsep}{6.5mm}
\setlength\bibsep{1pt}

\makeatletter
\newlength{\figrulesep}
\setlength{\figrulesep}{0.5\textfloatsep}

\newcommand{\topfigrule}{\vspace*{-1pt}%
	\noindent{\color{cream}\rule[-\figrulesep]{\columnwidth}{1.5pt}} }

\newcommand{\botfigrule}{\vspace*{-2pt}%
	\noindent{\color{cream}\rule[\figrulesep]{\columnwidth}{1.5pt}} }

\newcommand{\dblfigrule}{\vspace*{-1pt}%
	\noindent{\color{cream}\rule[-\figrulesep]{\textwidth}{1.5pt}} }

\makeatother

\twocolumn[
  \begin{@twocolumnfalse}
{\includegraphics[height=30pt]{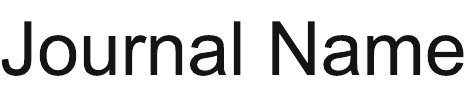}\hfill\raisebox{0pt}[0pt][0pt]{\includegraphics[height=55pt]{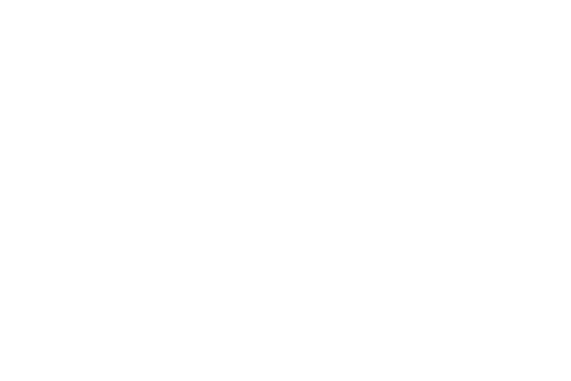}}\\[1ex]
\includegraphics[width=18.5cm]{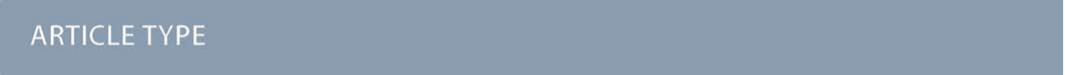}}\par
\vspace{1em}
\sffamily
\begin{tabular}{m{4.5cm} p{13.5cm} }

\includegraphics{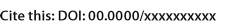} & \noindent\LARGE{\textbf{Striking the Right Balance of Encoding Electron Correlation in the Hamiltonian and the Wavefunction Ansatz}} \\
\vspace{0.3cm} & \vspace{0.3cm} \\

 & \noindent\large{Kalman Szenes\footnotemark[1], Maximilian M\"orchen\footnotemark[1], Paul Fischill\footnotemark[1], and Markus Reiher\footnotemark[1]\footnotemark[2]} \\

\includegraphics{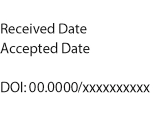} & \noindent\normalsize{
Multi-configurational electronic structure theory delivers the most versatile approximations to many-electron
				wavefunctions, flexible enough to deal with all sorts of
				transformations, ranging from electronic excitations,
				to open-shell molecules and chemical reactions. Multi-configurational models are therefore essential to establish universally
				applicable, predictive ab initio methods for chemistry.
				Here, we present a discussion of explicit correlation approaches
				which address the nagging problem of dealing with static and dynamic electron correlation in multi-configurational active-space
				approaches. We review the latest developments and then point to their key obstacles.
				Our discussion is supported by new data obtained with tensor network methods. We argue in favor of simple electrons-only correlator expressions that may allow one to define transcorrelated models in which the correlator does not bear a dependence on molecular structure.} \\
\end{tabular}

 \end{@twocolumnfalse} \vspace{0.6cm}

  ]

\renewcommand*\rmdefault{bch}\normalfont\upshape
\rmfamily
\section*{}
\vspace{-1cm}


\footnotetext[1]{Department of Chemistry and Applied Biosciences, ETH Z\"urich, Vladimir-Prelog-Weg 2, 8093 Z\"urich, Switzerland}

\footnotetext[2]{Email: mreiher@ethz.ch (corresponding author)}


\section{Introduction}
Closed-shell molecules can be described by a single reference determinant (the Hartree-Fock determinant) and single-reference coupled cluster (CC) theory then yields consistently high accuracy which can be increased in a systematic manner~\cite{helgaker_molecular_2000}.
Standard CC models~\cite{bartlettCoupledclusterTheoryQuantum2007,shavitt_many-body_2009} are available for reliable calculations of chemical accuracy, and higher accuracy can be achieved by increasing the excitation rank of the cluster operator.
Complicated electronic structures, such as those found in open-shell transition metal complexes and clusters, unsaturated molecules, and biradicals, require a different ansatz. In such systems, the Hartree-Fock determinant will not dominate the determinantal expansion of the electronic wavefunction, hence, single-reference approaches based on it are likely to break down.

Multi-configurational wavefunction methods usually start from (complete) active orbital spaces and deliver a qualitatively correct description of the valence shell of electronic states and are, therefore, the standard starting point for tackling challenging electronic structure problems.
However, this ansatz introduces by definition an imbalance in the description of electron correlation due to the resulting (somewhat arbitrary) separation of electron correlation into static and dynamic contributions within and outside the active orbital space chosen.
Therefore, a holy grail in electronic structure theory has been the accurate calculation of electronic energies in such composite approaches for static and dynamic electron correlation as they eventually produce unbalanced error contributions. 

Active space methods rely on the restriction of the orbital space in which a full configuration interaction (FCI) wavefunction is then constructed (the classic example is the complete active space (CAS)~\cite{roosCompleteActiveSpace1980, roos1980b} or full optimized reaction space~\cite{ruedenbergAreAtomsIntrinsic1982} ansatz).
However, the orbitals that are ignored in such a CAS-CI ansatz significantly contribute to the electronic energy, which cannot be ignored. 

To include such dynamic correlations~\cite{rocasanjuan_multiconfiguration_2012}, 
CC approaches are the natural choice to turn to.
However, multi-reference CC ans\"atze are plagued by various formal and practical limitations: the \textit{multiple-parentage} problem, intruder states~\cite{schucan1972, schucan1973,jankowski1992,paldus1993}, lack of orbital invariance, increased scaling with the active space size~\cite{hanrath2008}, or non-truncating cumulant expansion~\cite{yanai2006, yanai2007, evangelista2012a}.
Even though most of these drawbacks can be circumvented by the multi-reference formulation of configuration interaction (CI), this approach requires expensive higher-order density matrices and, due to its truncated CI expansion, suffers from size-consistency issues as does truncated single-reference CI~\cite{lyakh2012,evangelista2018}.
For an in-depth description of multi-reference methods, see Refs.~\citenum{lyakh2012, evangelista2018}.

As a result, various other approaches have been designed to assess dynamic correlation for multi-configurational methods; examples are, multi-reference perturbation theories~\cite{Andersson1992Jan,Angeli2001Jun,Lindh2020Nov}, short-range density functional theory~\cite{leiningerCombiningLongrangeConfiguration1997,Fromager2007Feb,Hedegard2015Jun,Hedegard2018Jun,Rodriguez-Jimenez2021Mar,Pernal2022Mar}, 
multi-configurational pair-density functional theory approaches \cite{LiManni2014Sep,Sharma2021Apr},
and multi-reference-driven single-reference CC 
models (see for example Refs.~\citenum{Oliphant1991,Piecuch_1993,Adamowicz2000_cascc1,Ivanov2000_cascc2,kinoshita2005,veis2016a,faulstich2019,vitale2020a,vitale2022}.
We note also that much work aimed to describe both static and dynamic correlation accurately in the CC framework alone;
see, for instance, CC approaches with internal and semi-internal excitations\cite{Piecuch_1993,piecuch1999}, method-of-moments CC\cite{piecuch2004,kowalski2000,kowalski2000a}, and CC(P;Q)\cite{shen2012,shen2012a,shen2012b,bauman2017,deustua2017}.
For further CC-based methods that aim to describe both static and dynamic correlation, see Ref.~\citenum{piecuch2010} and references therein.

In the end, multi-configurational methods require a combined approach, which treats dynamic correlation on a different footing, compromising the overall accuracy.
As a result, no multi-configurational approach is known to achieve a consistent and systematically improvable accuracy as the single-reference CC model does.
Therefore, it is highly desirable to establish a formal theory that can address and eventually solve this problem.
Here, we consider explicitly correlated ans\"atze as a possible solution to the problem.
Many ideas have already been put forward, and remarkable successes have been achieved (especially when considering developments in explicitly correlated CC models~\cite{klopper2006,ten-no2012,hattig2012,kong2012}).
However, no universal framework has emerged, and various conceptual problems remain open or, at least, require further discussion.
Therefore, we first prepare a general formal framework for these approaches.
Then, we highlight features of the various proposals made in the context of explicit correlation theory so far, discuss key challenges, and provide new data to substantiate these challenges.
After this analysis, we settle on basic principles that could be considered universally valid frameworks for explicit correlation approaches.

\section{Theory}

\subsection{General Theory}

Under the Born-Oppenheimer approximation, the most general expression for the electronic Hamiltonian $\hat{H}_\mathrm{el}$ may be written in first quantization for $N_\mathrm{el}$ electrons as
\begin{equation}
	\hat{H}_{\mathrm{el}} = \sum_i^{N_\mathrm{el}} \hat{h}(i) + \sum_{i < j}^{N_\mathrm{el}} \hat{g}(i,j).
\end{equation}
The terms $\hat{h}(i)$ and $\hat{g}(i, j)$ correspond to the one- and two-electron operators, respectively.
The operator $\hat{h}(i)$ comprises the electronic kinetic energy operator, which may be chosen according to Dirac or in its non-relativistic limit according to Schr\"odinger for infinite speed of light\cite{reiher_relativistic_2014},
and the external potential created by the nuclei.
For the sake of simplicity, we omitted the nucleus-nucleus interaction, which contributes a constant term in the Born-Oppenheimer approximation and can be added at any later stage.

The two-electron operator $\hat{g}(i, j)$ describes the electron-electron interaction.
Its leading term is the electrostatic Coulomb interaction of the electrons.
In the non-relativistic limit of Schr\"odinger quantum mechanics, this will be the only contribution.
Magnetic and retardation effects contribute to higher order in the inverse speed of light $1/c$. They can be taken into account to order $1/c^2$ by Gaunt and Breit operators, containing the inverse of the particles' distance (as in the case of the Coulomb interaction) and even up to the inverse distance cubed in the case of the second term of the Breit operator.
We note that interactions involving the nuclei are also affected by magnetic and retardation effects, which are usually negligible.

Whichever of these models is chosen for $\hat{H}_{\mathrm{el}}$, already the Coulomb potential generates singularities in the operator at the coalescence of two or more particles.
This property leads to well-known coalescence conditions~\cite{Kato1957Jan,pack_cusp_1966,tewSecondOrderCoalescence2008} in the exact wavefunction which are notoriously challenging to describe by conventional orbital-based wave-function ans\"atze.
This is due to the need for high angular momentum basis functions in order to accurately describe the linear behavior of the wavefunction in the neighborhood of the point of coalescence of two particles~\cite{Schwartz1962May}.
For the purpose of this article, we will restrict ourselves to the standard model, the electrostatic Coulomb interaction.

The central problem in electronic structure theory is obtaining the solutions to the electronic-structure eigenvalue problem for some electronic state $\psi_i$
\begin{equation}\label{eq:eig-prob}
  \hat{H}_{\mathrm{el}} \psi_i = E_i \psi_i.
\end{equation}
The solutions to the previous equation possess a factorized representation~\cite{fournaisSharpRegularityResults2005} 
\begin{equation}\label{eq:jastrow-slater}
	\psi_i = F \phi_i
\end{equation}
where $F$ is a universal, state-independent,
function, referred to as the \textit{correlator}, such that $\phi_i$ possesses continuous first-order derivatives.
Recall that $\psi_i$ is known to contain cusps at the point of particle coalescence while the function $\phi_i$ remains smooth.
Hence, the effect of the correlator $F$ is to remove the cusps present in the original wavefunction $\psi_i$.
The hope is that, by using the factorized representation as an ansatz, the problematic short-range correlation may be accounted for by the function $F$, and the remaining long-range correlation may be efficiently approximated by $\phi_i$.

\subsection{Cusp Conditions}\label{sec:cusp}
In the seminal work by Kato~\cite{Kato1957Jan}, the analytic behavior of the exact electronic wavefunction in the vicinity of particle coalescence was derived, leading to the cusp conditions that bear his name.
Subsequently, his results were extended~\cite{pack_cusp_1966} by relaxing the fixed-nuclei assumption and deriving coalescence conditions for wavefunctions that contain a node at the point of coalescence.
Later, Fournais and collaborators~\cite{fournaisSharpRegularityResults2005} introduced three-particle (electron-electron-nuclear) coalescence conditions and proved not only that the universal state independent factor $F$ from Eq.~(\ref{eq:jastrow-slater}) exists such that $\phi_i$ has continuous first-order derivatives but also derived its explicit form.
Additionally, Tew~\cite{tewSecondOrderCoalescence2008} derived higher-order coalescence conditions in the inter-particle distance and showed that these terms introduce system and state-dependent coefficients.

The cusp conditions may be classified by the type and number of particles coalescing:
\begin{enumerate}
	\item Nucleus-electron coalescence:

	      Under the Born-Oppenheimer approximation,
	      the position of the nuclei is fixed.
	      Then, for point-like nuclei, the (non-relativistic) electronic wavefunction is expected to contain a cusp at the nuclear position, which may be easily verified by plotting the wavefunction, as done in standard textbooks~\cite{helgaker_molecular_2000}.

	      Precisely describing this cusp in the wavefunction is crucial for computing the total energy to high accuracy.
	      However, in the Born-Oppenheimer approximation, errors in the electronic wavefunction at the nuclei are atomically conserved across the potential energy surface as long as one is interested in the valence-shell properties of atoms, molecules, and materials.
	      Therefore, these errors can be expected to cancel when properties, such as energy differences or derivatives, are considered. As an example, we may refer to comparisons of energies obtained for point-like and finite-size nuclei.
       Both types of nuclear charge distribution models affect the wavefunction in profoundly different ways at electron-nucleus coalescence points~\cite{reiher_relativistic_2014},
          which affects, in turn, total energies significantly but leaves relative energies essentially unchanged (see, for example, Ref.~\citenum{Andrae2000Apr}).

       	      We emphasize that the Coulomb singularities arise solely from point particles.
        By considering finite-size nuclear charge distribution models~\cite{Andrae2000Oct}, which is particularly important for large nuclei, the electron-nucleus cusps may be entirely eliminated (although a steep increase of the wavefunction, and hence of the electron density, in the vicinity of atomic nuclei will still be observed).
       
	      It is for this reason that we ignore electron-nucleus cusps in this work.
	      This assumption will not hold, however, if we consider properties that probe electron density close to the nucleus, as in the case of the isomer shift of M\"ossbauer spectroscopy~\cite{proppeReliableEstimationPrediction2017}.
	      In addition, by assuming the nuclei to be static, the location of these cusps is known a priori, and therefore, atom-centered one-electron basis functions may be employed that explicitly satisfy the cusp condition.
	      This is the case for Slater orbitals~\cite{Slater1928Mar}, and techniques have been devised that also augment existing Gaussian basis sets with this property~\cite{maSchemeAddingElectron2005}.

	\item Electron-electron coalescence:

	      Accurately describing the coalescence of two electrons is at the heart of capturing the short-range dynamic correlation of electrons.
	      While the electron-nucleus coalescence can be simply visualized, the electron-electron cusp is more subtle since both of the particles are considered dynamic (i.e., their coordinates are both variables in electronic structure theory) and thus, any position in space may be chosen for the coordinate of one electron and used to study the behavior of the wavefunction as another electron is brought nearer.
        In practice, the electron-electron coalescence conditions manifest themselves in the evaluation of the electronic repulsion integrals.

        The electronic coalescence conditions are generally classified into either singlet or triplet, depending on whether the spin function of the considered pair of electrons is symmetric or not with respect to the permutation of the two electrons.
	      For singlet electron pairs, which have antisymmetric spin functions, the cusp condition enforces the following form on the exact wavefunction~\cite{Kato1957Jan} 
	      \begin{equation}
		      \psi\ =\ \left(1\ +\ \frac{1}{2}r\right)\psi\left(r = 0\right) +  \mathcal{O}\left(r^{2}\right),
	      \end{equation}
	      where $r$ is the inter-electronic distance.

	      For triplet pairs, the exact wavefunction vanishes at the point of coalescence.
	      Moreover, it possesses continuous first-order derivatives, and the cusp only appears in the second-order derivative\cite{pack_cusp_1966}.
	      The exact wavefunction, therefore, obeys a different condition
	      \begin{equation}
		      \psi = \mathbf{r}\cdot\frac{\partial\psi}{\partial{\mathbf{r}}}\biggl|_{r=0}\left(1 + \frac{1}{4}r\right) + \mathcal{O}\left(r^{3}\right).
	      \end{equation}
	      A consequence of this property is that certain states, such as high-spin open-shell systems, do not possess electronic cusps in the wavefunction (only in their derivative).

	\item Three or more particle coalescence:
 
	      In principle, the behavior of the wavefunction at the point of coalescence of multiple particles can be derived.
        However, specific findings~\cite{fournaisSharpRegularityResults2005,tewSecondOrderCoalescence2008} have only been attained for three particle coalescence points, specifically for electron-electron-nucleus coalescence.
        While numerical results suggested that the consideration of these terms can improve the accuracy of the model~\cite{Cohen2019Aug,Ammar2023Aug}, we argue that the small probability of three particles simultaneously coinciding at the same point makes this effect play only a minor role in the description of the wavefunction.

\end{enumerate}

\subsection{Cusp Conditions and Correlators}
Next, it is key to consider how the coalescence conditions affect the choice of the correlator $F$ in general.
For this purpose, we revisit in more detail the analytic form of $F$ derived by Fournais et al.~\cite{fournaisSharpRegularityResults2005}.
Assuming an exponential form $F = e^{\tau}$ for the ansatz in Eq. (\ref{eq:analytic-corr}),
Fournais and collaborators derived that $\tau$ must be composed of three types of terms~\cite{tewSecondOrderCoalescence2008}
\begin{equation}\label{eq:analytic-corr}
  \tau = - \sum_I^{N_\mathrm{nuc}} \sum_i^{N_\mathrm{el}} Z_I r_{iI} + \frac{1}{2} \sum_{i < j}^{N_\mathrm{el}} r_{ij} + \frac{2 - \pi}{6 \pi} \sum_I^{N_\mathrm{nuc}} \sum_{i < j}^{N_\mathrm{el}} Z_I \mathbf{r}_{iI} \cdot \mathbf{r}_{jI} \ln(r_{iI}^2 + r_{jI}^2)
\end{equation}
where the upper case letters $I$ and lower case letters $i$ correspond to nuclear and electronic indices, respectively, and $Z_{I}$ represents the nuclear charge of particle $I$
(note the prefactor in the definition of the kinetic energy operator in Ref. ~\citenum{fournaisSharpRegularityResults2005}).
The first and second terms in Eq.~(\ref{eq:analytic-corr}) account for the nucleus-electron and electron-electron cusps, respectively, while the last term enforces the three-particle coalescence condition, namely the electron-electron-nucleus coalescence condition.

In light of the discussion so far, we argue for omitting the first term, i.e., the one taking care of all electron-nucleus cusps, since we maintain that, although its effect on the total electronic energy may be non-negligible, it introduces an atomically conserved error that drops out for relative electronic energies dominated by changes in the valence region of atoms and molecules. 
For similar reasons, although the three-particle term has been shown to have also a visible effect on the total electronic energy (see Ref.~\citenum{Myers1991Nov} for numerical data and also for a historical perspective on this contribution), we argue that this effect is not likely to be strong for relative properties dominated by valence-shell contributions.
Therefore, we contend that its omission will also benefit from systematic error cancellation.
We note, however, that correlators that include terms of the first type have been employed recently in the context of transcorrelated multi-reference theories~\cite{Cohen2019Aug,Ammar2023Aug} (see also below).

We highlight that Eq.~(\ref{eq:analytic-corr}) inherently introduces system-dependent quantities into the correlator through the presence of the nuclear positions and charges.
However, neglecting the first and third terms removes all system dependence of the correlator and requires the remainder of the wavefunction, that is, $\phi_i$, of Eq.~(\ref{eq:analytic-corr})
to take care of the system- and state-dependence. Note that Tew~\cite{tewSecondOrderCoalescence2008}
explicitly derived the system- and state-dependence for a spherical model of a particle's wavefunction around a coalescence point in the form of higher-order expansion coefficients in the partial wave expansion. However, these results are not directly transferable to the product ansatz of Eq.~(\ref{eq:analytic-corr}) 
that we exploit in this work.

For these reasons, we here advocate for the use of a simple correlation factor, which may be taken to be universal, state- and system-independent, in order to define a comparatively simple yet (for valence-shell dominated relative energies and properties) accurate, explicitly correlated electronic structure method.
Hence, the remaining term responsible for the electron-electron cusp is considered the only essential term in Eq.~(\ref{eq:analytic-corr}) for an electronic structure model defined within the Born-Oppenheimer approximation.

\subsection{Energy Evaluation with Correlators}\label{sec:eeval-corr}

It is well established that the description of the cusp is the dominating cause of the slow convergence of standard ab initio methods with increasing size of one-electron (orbital) basis sets~\cite{Schwartz1962May}.
The fact that the wavefunctions may be factorized as in Eq.~(\ref{eq:jastrow-slater}), with the function $\phi$ being smooth and devoid of cusps, lends it to easier treatment by conventional one-particle Gaussian basis sets.

By choosing the function $F$ such that the cusp conditions are obeyed, Eq.~(\ref{eq:jastrow-slater}) may be employed as an ansatz for the wavefunction.
Placing this ansatz into the eigenvalue problem in Eq.~(\ref{eq:eig-prob}) yields
\begin{equation}
  \hat{H}_{\mathrm{el}} F \ket{\phi} = E F \ket{\phi}.
\end{equation}
where we assume a fixed correlator $F$ and a determinantal expansion of $\phi$, denoted by the Dirac notation (note that we dropped the state index for the sake of brevity).
This equation may be solved for $E$ by multiplying by $\bra{F \phi}$ on the left and isolating $E$ to produce the familiar Rayleigh quotient
\begin{equation}\label{eq:variational}
  E = \frac{\bra{\phi}F^{\dagger} \hat{H}_{\mathrm{el}} F \ket{\phi}}{\bra{\phi}F^{\dagger}F\ket{\phi}}.
\end{equation}
The parameters in the $\ket{\phi}$ may be optimized variationally, and the obtained energy will always yield an upper bound on the exact value.
Unfortunately, this expression requires the evaluation of $N_{el}$-electron integrals~\cite{Francis1969Apr}, which makes it intractable for even the smallest of systems.

An alternative expression for the energy may be obtained through projection onto $\bra{\phi} F^{-1}$
\begin{equation}\label{eq:sim-transform}
  E = \frac{\bra{\phi} F^{-1} \hat{H}_{\mathrm{el}} F \ket{\phi}}{\braket{\phi|\phi}}
\end{equation}
which only requires the evaluation of up to three-electron integrals~\cite{Francis1969Apr}.
The operator $\bar{H} = F^{-1} \hat{H}_{\mathrm{el}} F$ corresponds to a similarity transformation of the operator $\hat{H}_{\mathrm{el}}$ by the correlator $F$.
This transformation has the property of preserving the spectrum of the original operator, as it can be interpreted as a change of basis.
However, the operator $\bar{H}$ is no longer necessarily Hermitian, and thus Eq.~(\ref{eq:sim-transform}) is not bounded from below~\cite{Lowdin1988Jan}; preventing the use of variational techniques for optimizing $\ket{\phi}$.

The two equations for the energy seem quite similar, and one might ask how they are related.
A first observation is that, if the operator $F$ is unitary (i.e., $F^{-1} = F^{\dagger}$), the two expressions actually coincide.
Moreover, a property of unitary transformations is that they preserve the norm of vectors, and therefore, by assuming a normalized $\ket{\phi}$, the term in the denominator vanishes
\begin{equation}
  E = \bra{\phi} F^{\dagger} \hat{H}_{\mathrm{el}} F \ket{\phi}.
\end{equation}

As a comment, we recall that certain properties of operators no longer hold once these operators are projected on a finite basis, as done in the second quantization formalism.
For instance, projecting two operators into a finite basis and taking their matrix product is not necessarily equal to the product of the two operators projected into the same basis~\cite{helgaker_molecular_2000}.
This has the consequence that the similarity transformed operator in Eq.~(\ref{eq:sim-transform}) is only guaranteed to have the same spectrum as the original operator in the limit of a complete basis.

\subsection{Similarity Transformation in Coupled Cluster Theory}
A similarity transformation of the Hamiltonian is also at the heart of CC theory, which employs an exponential ansatz for the wavefunction
\begin{equation}
	\ket{\psi} = e^{\hat{T}} \ket{\phi}
\end{equation}
with a reference wavefunction $\ket{\phi}$.
The cluster operator is given by
\begin{equation}
	\hat{T} = \hat{T}_1 + \hat{T}_2 + \ldots
\end{equation}
where $\hat{T}_i$ are the standard $i$-fold excitation operators parameterized with unknown amplitudes $t^i_a, t^{ij}_{ab}, \ldots$.
These amplitudes are determined iteratively by solving the CC equations, which project the Schr\"odinger equation onto the excited determinants $\bra{\psi^\mu}$
\begin{equation}\label{eq:cc-t}
    0 = \braket{\psi^\mu | e^{-\hat{T}} \hat{H}_{\mathrm{el}} e^{\hat{T}} | \phi}.
\end{equation}
For convenience, we denote the similarity transformed operator by $\tilde{H} = e^{-\hat{T}} \hat{H}_{\mathrm{el}} e^{\hat{T}}$.
Once the amplitudes have converged, the electronic energy may be computed through
\begin{equation}\label{eq:cc-e}
  E_{\mathrm{CC}} = \braket{\phi | \tilde{H} | \phi}.
\end{equation}
For simplicity, consider a two-electron system for which the CCSD ($\hat{T} = \hat{T}_1 + \hat{T}_2$) method delivers the exact solution.
Eq.~(\ref{eq:cc-t}) and Eq.~(\ref{eq:cc-e}) may be combined into a matrix equation which, for converged amplitudes, has the following structure
\begin{equation}
  \tilde{H} = \begin{pmatrix}
    E_{\mathrm{CC}}  & \tilde{H}_{0S}  & \tilde{H}_{0D} \\
        0 & \tilde{H}_{SS}  & \tilde{H}_{SD} \\
        0 & \tilde{H}_{DS}  & \tilde{H}_{DD}
        \end{pmatrix}.
\end{equation}
The first column corresponds to the CC energy and amplitude equations, while the other elements correspond to the remaining transformed matrix elements of $\hat{H}_{\mathrm{el}}$.
Note that the structure of the first column guarantees that the reference wavefunction is an eigenfunction of $\tilde{H}$.
Hence, CC can be interpreted as optimizing the cluster amplitudes such that the reference wavefunction, corresponding to the first column, becomes an eigenvector of the similarity-transformed Hamiltonian $\tilde{H}$.
By recalling that $\ket{\phi}$ is usually taken as the Hartree-Fock determinant, which only accounts for mean-field interactions, the similarity transformation has effectively transferred all electronic correlation directly into the operator $\tilde{H}$.

In CC theory, the amplitudes may also be obtained variationally~\cite{Szalay1995Jul} through the use of a symmetric expectation value
\begin{equation}\label{eq:var-cc}
    E_{\mathrm{CC}} = \frac{\braket{\phi | (e^{\hat{T}})^\dagger \hat{H}_{\mathrm{el}} e^{\hat{T}} | \phi }}{\braket{\phi | (e^{\hat{T}})^\dagger e^{\hat{T}} | \phi }}.
\end{equation}
This expression, however, is not used in practice since it does not benefit from the natural truncation of the Taylor expansion of the transformed operator and therefore, the evaluation of Eq.~(\ref{eq:var-cc}) requires the computation of matrix elements between determinants of arbitrary excitation rank; making the expression exponentially difficult to evaluate, even for truncated $\hat{T}$.
A noteworthy exception is unitary-CC~\cite{Kutzelnigg1991Jul} (UCC) in quantum computing ~\cite{Romero2018Oct} since unitary gates may be directly implemented on quantum hardware.
In UCC, a similar expression to Eq. (\ref{eq:var-cc}) is used, where the cluster operator $\hat{T}$ is replaced by an anti-Hermitian operator $\hat{\sigma} = \hat{T} - \hat{T}^\dagger$ which results in a unitary similarity transformation; preserving the Hermiticity of the original operator.

The inadequacies of Eq.~(\ref{eq:var-cc}) seem in complete analogy with the ones discussed in Sec.~(\ref{sec:eeval-corr}) with the correlator $F$ being given by the cluster operator $e^{\hat{T}}$.
A crucial difference is, however, that in the case of CC theory, the excitation operators $\hat{T}$ are assumed to be restricted to creating excited determinants in the same Hilbert space as the one spanned by the one-particle basis.
Therefore, $E_{\mathrm{CC}}$ can, at best, converge to the FCI solution contained in the finite Hilbert space.
This restriction is generally not true for an arbitrary correlator $F$ defined in real space.
In particular, when the correlator is expressed in real space, the excitations are inevitably produced in orbitals that are not contained in the one-particle basis, and hence, the computed energies may even improve upon the FCI solution.

While it is sometimes stated that the non-variational nature of the standard CC energy is due to the similarity transformation only being approximate~\cite{crawford_introduction_2000} -- mirroring the conclusions from the last paragraph of Sec.~(\ref{sec:eeval-corr}) -- in reality the similarity transformation is always exact, in the sense that no matter what amplitudes are inserted into the cluster operator, the spectra of the matrix representation of $\hat{H}_{\mathrm{el}}$ and $\tilde{H}$ always coincide.
The non-variationality of the scheme actually occurs due to the eigenvalue of $\tilde{H}$ only being calculated approximately.
For instance, in a four-electron system, if Eq.~(\ref{eq:cc-t}) is only solved using CCSDT, the resulting matrix representation would have the form
\begin{equation}
  \tilde{H} = \begin{pmatrix}
    E_{\mathrm{CCSDT}}  & \tilde{H}_{0S}  & \tilde{H}_{0D} & \tilde{H}_{0T} & \tilde{H}_{0Q} \\
        0 & \tilde{H}_{SS}  & \tilde{H}_{SD} & \tilde{H}_{ST} & \tilde{H}_{SQ} \\
        0 & \tilde{H}_{DS}  & \tilde{H}_{DD} & \tilde{H}_{DT} & \tilde{H}_{DQ} \\
        0 & \tilde{H}_{TS}  & \tilde{H}_{TD} & \tilde{H}_{TT} & \tilde{H}_{TQ} \\
        \tilde{H}_{Q0} & \tilde{H}_{QS}  & \tilde{H}_{QD} & \tilde{H}_{QT} & \tilde{H}_{QQ} \\
        \end{pmatrix}.
\end{equation}
In this case, the reference wavefunction is only an approximate eigenvector of $\tilde{H}$, and if the eigenvector was computed exactly by also projecting on the quadruply excited determinants, the energy would be equal to the energy computed from $\hat{H}$.

\subsection{Choice of the correlator}
While the goal is to select the correlator $F$ such that the remaining function $\phi$ is smooth, there remains significant freedom in the choice of its explicit form.
The remainder of this section is dedicated to an overview of the general class of correlators that have proved useful when applied to chemical systems.

\subsubsection{General Class of Correlators}
Historically, the first generally applicable explicitly correlated methods are the so-called R12 methods~\cite{Kutzelnigg1985Dec,Klopper1987Feb,klopper2006}.
These methods are defined by explicit dependence of the wavefunction on inter-electronic distance $r_{ij}$ through the multiplication of linear $r_{ij}$ terms to standard wavefunction ans\"atze.
Combined with their clever use of resolution of identities to compute the arising non-standard integrals, these methods became computationally feasible and thus applicable even to large systems~\cite{Klopper1996Nov}.

Subsequently, the class of R12 methods has been extended to accommodate correlators depending non-linearly on $r_{ij}$, which are commonly referred to as F12 methods~\cite{mayExplicitlyCorrelatedSecond2004}.
Examples of correlators used in F12 methods include Gaussian-type Geminals~\cite{perssonAccurateQuantumchemicalCalculations1996} $e^{-\gamma r_{ij}^2}$ and Slater-type Geminals~\cite{ten-no_initiation_2004-2} $e^{-\gamma r_{ij}}$.
Most of these correlators contain a number of free parameters which must be adjusted for the target system.
Some contain a single tunable parameter~\cite{ginerNewFormTranscorrelated2021} while others are composed of a handful of parameters~\cite{ten-noFeasibleTranscorrelatedMethod2000} whose optimal values may be deduced based on first principles.
In the context of Monte Carlo methods~\cite{schmidt_correlated_1990,Umezawa2003Nov,Drummond2005Aug,Haupt2023Jun}, flexible correlators are often used, which contain dozens of free parameters that are optimized in a black-box fashion using variance minimization techniques~\cite{handyMinimizationVarianceTranscorrelated1971}.
A comparison of the performance of a large set of commonly used correlators can be found in Refs.~\citenum{johnson_explicit_2017,tew_new_2005}.

\subsubsection{Gutzwiller Correlator}
A widely used correlator in the context of spin models and in solid state physics is the Gutzwiller correlator\cite{gutzwiller1963}. 
In contrast to explicit correlation, which operates in real space, the Gutzwiller correlator operates in Hilbert space by destabilizing doubly occupied states in the Hartree-Fock determinant based on a variational parameter, which can increase the convergence of correlated methods such as DMRG~\cite{baiardi_transcorrelated_2020} and FCIQMC~\cite{dobrautz2019}.
Different approaches exist to apply Gutzwiller-like correlators to molecular systems~\cite{yao2014,dong2020}, as well as approaches based on similarity transformation~\cite{tsuneyuki2008, neuscamman2011, wahlen_strothman2015} of the Hamiltonian with the correlator.
However, since this correlator acts in Hilbert space, the correlated and uncorrelated wavefunctions converge to the same FCI energy.

\subsubsection{Jastrow Factor}
For the purpose of this study, we only consider correlators $F$ that are parameterized by an exponential $F = e^\tau$, referred to as a Jastrow factor~\cite{jastrow_many-body_1955}.
The exponential parameterization of the correlator automatically guarantees size-consistency of the electronic energies, which is an important requirement for accurate results in quantum chemistry.
As argued in Sec.~(\ref{sec:cusp}), we limit the function $\tau$ to pair-wise electronic terms in $r_{ij}$.
These terms must be symmetric with respect to the permutation of two electrons to fulfill the wavefunction's antisymmetry properties, which is already satisfied by the determinantal expansion $\ket{\phi}$.

\section{Transcorrelation}
In the transcorrelated method~\cite{Francis1969Apr,handyEnergiesExpectationValues1969}, the conventional electronic Hamiltonian $\hat{H}_{\mathrm{el}}$ is similarity transformed by the correlation factor $e^{\tau}$
\begin{equation}
    \hat{H}_{\mathrm{tc}} = e^{-\tau} \hat{H}_{\mathrm{el}} e^{\tau}.
\end{equation}
yielding the transcorrelated Hamiltonian $\hat{H}_{\mathrm{tc}}$.
The correlation factor is chosen such that the product $e^{\tau}\ket{\phi}$ satisfies the cusp conditions. 
The operator $\hat{H}_{\mathrm{tc}}$ may be expanded using the Baker--Campbell--Hausdorff formula which truncates naturally after the second nested commutator~\cite{Francis1969Apr}
\begin{equation}
	\hat{H}_{\mathrm{tc}} = e^{-\tau} \hat{H}_{\mathrm{el}} e^{-\tau} = \hat{H}_{\mathrm{el}} + [\hat{H}_{\mathrm{el}}, \tau] + \frac{1}{2}[[\hat{H}_{\mathrm{el}}, \tau], \tau],
\end{equation}
due to the term $[[H_{el}, \tau], \tau]$ being a multiplicative factor (i.e., it does not contain any differential operators) and thus all higher-order commutators in the expansion vanish (see Ref.~\citenum{lee_studies_2023} for a modern derivation of this result).
The two additional terms in $\hat{H}_{\mathrm{tc}}$ make its treatment significantly more complicated than $\hat{H}_{\mathrm{el}}$.

First, while the second nested commutator preserves the Hermiticity of the original operator, the first one introduces a non-Hermitian contribution, which prevents the use of optimization techniques relying on the variational principle.
This is because, for non-Hermitian operators, the left and right eigenvectors do not generally coincide, and thus, the conventional Rayleigh quotient for computing eigenvectors no longer applies.
This issue may, however, be tackled by employing a biorthogonal approach, which allows for distinct left and right eigenvectors and has successfully been applied to the transcorrelated method~\cite{boys_calculation_1970,Fimple1976Jul,Hino2001Nov,ammar_transcorrelated_2023-1,lee_studies_2023}.
Moreover, this approach provides a framework for performing orbital optimization within the transcorrelated method~\cite{Ammar2023Aug,Kats2024Apr}.
Efforts have also been made to apply the variational principle directly to the transcorrelated method by disregarding the non-Hermitian terms~\cite{luo_variational_2010,Luo2011Jul} in $\hat{H}_\mathrm{tc}$.

Second, the nested commutator introduces a three-body operator, which significantly increases the computational cost of the transcorrelated method compared to the conventional two-body electronic Hamiltonian.

Finally, both additional terms introduce non-standard integrals, which need to be evaluated, as is the case for F12 methods.
Approaches for computing these integrals include grid-based methods~\cite{boysFirstSolutionLiH1969,Cohen2019Aug}, density-fitting techniques~\cite{ten-noDensityFittingDecomposition2003,baiardi_explicitly_2022} and Monte Carlo methods~\cite{Umezawa2005Jun}.

The addition of the three-body operator in $\hat{H}_{\mathrm{tc}}$ poses a significant challenge for the transcorrelated approach due to the steep increase in required storage for the generated integrals as well as the computational cost of working with them.
For instance, in the case of DMRG, the three-body contribution increases the computational cost of the tensor contractions by two orders of magnitude~\cite{baiardi_explicitly_2022}.
Recently, a promising remedy~\cite{Liao2021Jul,Schraivogel2021Nov,Christlmaier2023Jul} for taming the expensive three-body operator has emerged.
It is based on the normal-ordering of the operators in $\hat{H}_{\mathrm{tc}}$ with respect to a reference state, usually chosen as the Hartree-Fock solution following the particle-hole formalism.
This allows one to include the mean-field, one- and two-body contribution from the three-body operator and leaves a "pure" three-body contribution, which is presumed to be small so that it may be neglected.
The validity of this approximation has been demonstrated~\cite{Christlmaier2023Jul} on a set of atoms and small molecules contained in the HEAT~\cite{Tajti2004Dec} benchmark dataset.

The transcorrelated method has a number of advantages over other explicitly correlated methods.
First, by using a projective technique to solve for the energy, the cusp conditions may be satisfied while limiting the required integrals to at most three-electron ones.
Second, by directly folding the correlation into the Hamiltonian, through the similarity transformation, and optimizing the wavefunction with this transformed operator, the correlation captured by the determinantal expansion is guaranteed to not overlap with the one already accounted for by the correlator~\cite{Cohen2019Aug}.
Indeed, to ensure this property for R12/F12 methods, orthogonality conditions~\cite{Szalewicz1982Sep,Werner2007Apr} need to be enforced in order to guarantee that the correlator generates excitations outside the Hilbert space spanned by the finite basis.
This formalism is quite cumbersome and usually limits the flexibility of correlators to simple functions~\cite{Cohen2019Aug}.

\subsection{Transcorrelated Density Matrix Renormalization Group (tcDMRG)}
Recently, the transcorrelated method has been adapted to the Density Matrix Renormalization Group (DMRG)~\cite{White1993Oct} first through the imaginary-time evolution formalism~\cite{baiardi_transcorrelated_2020,baiardi_explicitly_2022} and subsequently using the time-independent optimization scheme~\cite{Liao2023Mar}.
This section reviews key aspects of the imaginary-time evolution variant of tcDMRG developed in our group~\cite{baiardi_transcorrelated_2020,baiardi_explicitly_2022}.

In the matrix product operator (MPO) formalism~\cite{mcculloch_density-matrix_2007,Keller2015Dec} 
of DMRG, a matrix product state (MPS) represents the ansatz for the wavefunction
\begin{equation}
  \ket{\phi_{\mathrm{MPS}}} = \sum_{\boldsymbol{\sigma}}^{L}\sum_{\boldsymbol{\alpha}} M_{1, \alpha_1}^{\sigma_1} M_{\alpha_1, \alpha_2}^{\sigma_2} \ldots M_{\alpha_{L-1}, 1}^{\sigma_L} \ket{\boldsymbol{\sigma}}
\end{equation}
where $L$ corresponds to the number of molecular orbitals and $\sigma_i = \{0, \uparrow, \downarrow, \uparrow\downarrow \}$ denotes to the electronic occupation of orbital $i$.
Note that the first and last of these indices have an extent of only one in order to ensure the correct dimensionality of the contracted product of tensors.

The maximum values of the auxiliary indices $\alpha_i$ introduced in the factorization are given by the bond dimension. The bond dimension is the key parameter for determining the accuracy and cost of the DMRG algorithm~\cite{Keller2014Apr}.
For the exact FCI solution, the extent of the indices $\alpha_i$ must be allowed to grow exponentially along the chain.
In practice, the bond dimension is fixed to a maximal value for the duration of the optimization, and the MPS tensors are truncated to this value.
However, the validity of a chosen bond dimension can be probed rigorously by inspection of the singular value decompositions inherent to the DMRG algorithm and by systematic extrapolation to infinite bond dimension~\cite{Hubig2018Jan}.

Similar to the wavefunction, the Hamiltonian is factorized into an MPO
\begin{equation}
	\hat{H} = \sum_{\boldsymbol{\sigma}}^L\sum_{\boldsymbol{\beta}} W_{1, \beta_1}^{\sigma_1, \sigma_1'} W_{\beta_1, \beta_2}^{\sigma_2, \sigma_2'} \ldots W_{\beta_{L-1}, 1}^{\sigma_L, \sigma_L'} \ket{\mathbf{\sigma}}\bra{\mathbf{\sigma}}
\end{equation}
which, in the case of tcDMRG, corresponds to $\hat{H}_{\mathrm{tc}}$.
While the MPS and MPO factorizations do seem analogous, a crucial difference is that the MPS is an approximation to the true FCI wavefunction if the bond dimension is truncated, whereas the MPO comprises an exact representation of the Hamiltonian.

In tcDMRG, due to the non-Hermiticity of $\hat{H}_{\mathrm{tc}}$, the variational principle no longer holds so that the conventional DMRG optimization of the entries in the MPS is no longer directly applicable.
Hence, in order to optimize the MPS, we rely on the imaginary-time tangent-space formulation of DMRG (iTD-DMRG)~\cite{haegemanUnifyingTimeEvolution2016,Lubich2015Mar,baiardi_large-scale_2019} in which the imaginary-time evolution of the MPS is performed using the time-dependent Schr\"odinger equation in the manifold of MPS with fixed bond dimension $m$
\begin{equation}
  \frac{\mathrm{d}\ket{\phi_{\mathrm{MPS}}(t)}}{\mathrm{d}t}=-\mathcal{P}_{\mathrm{m}}\hat{H}_{\mathrm{tc}}\ket{\phi_{\mathrm{MPS}}(t)}
\end{equation}
The operator $\mathcal{P}_{\mathrm{m}}$ projects the product $\hat{H}_{\mathrm{tc}} \ket{\phi_{\mathrm{MPS}}(t)}$ onto this manifold.
At $t \to \infty$, the solution converges to the optimal approximation of the ground state solution in the manifold.
In our implementation, this propagation is performed using a second-order Trotterization scheme.

Following our original work on tcDMRG~\cite{baiardi_transcorrelated_2020,baiardi_explicitly_2022}, we employ a correlation factor
\begin{equation}
    \tau = \frac{1}{2} \sum_{i < j} r_{ij}e^{-\gamma r_{ij}}
\end{equation}
with a single tunable parameter $\gamma$ in an exponential that plays the role of a damping function.
In the vicinity of particle coalescence, this correlator satisfies the cusp condition~\cite{tew_new_2005} and, by virtue of the damping function $\exp{(-\gamma r_{ij})}$, it diminishes at large particle distances to not interfere with long-range dynamic correlations.
Note that in the limit of $\gamma \to \infty$ the correlator vanishes so that $\hat{H}_{\mathrm{tc}}$ reduces to the conventional Hamiltonian $\hat{H}_{\mathrm{el}}$.

Although the damping function is a necessary ingredient in this ansatz, it introduces the somewhat arbitrary parameter $\gamma$ that prevents this specific tcDMRG approach from being a well-defined electronic structure model. For this reason, the
effect of this parameter on the tcDMRG energy will be investigated in this work.

\section{Results}

\subsection{Lithium Hydride}
As an example, we chose the Lithium Hydride molecule for which $\hat{H}_{\mathrm{tc}}$ can be treated with the full three-body operator with our implementation of tcDMRG.
These results can be used to validate the normal-ordered approximation of the three-body operator, in which the normal-ordered three-body operator with respect to the Hartree-Fock determinant is neglected.
Subsequently, we calculated the potential energy curve for LiH in the normal-ordered approximation and assessed the effect of the value of the $\gamma$ in the damping function of the correlator.

\subsection{Computational Methodology}
The tcDMRG calculations were performed with a development version of our DMRG program QCMaquis~\cite{Keller2015Dec}.
First, 10 initial sweeps (back-and-forth) across the orbital lattice were conducted using the time-independent variational DMRG algorithm with the conventional electronic Hamiltonian, assuming that this would lead to a good starting guess for the transcorrelated calculations.
Then, the wavefunction was optimized with the transcorrelated Hamiltonian using iTD-DMRG~\cite{baiardi_large-scale_2019} until energy convergence within a threshold of $10^{-10}$ Hartree.
Unless stated otherwise, a bond dimension of $m = 100$ was chosen, which we found to be sufficient to converge DMRG to the FCI solution for our model system (see, for instance, Table~\ref{tab:Eeq-De}).
In all simulations, the two-site optimization variant of DMRG was used.
The calculations were performed using the Dunning family of basis sets~\cite{Dunning1989Jan} cc-pVXZ (X = D, T, Q, 5), with the corresponding RIFIT basis~\cite{Weigend1998Sep} used in the density-fitting of the two-electron integrals, while the uncontracted cc-pV5Z basis set was used for the evaluation of the three-body integrals.
These integrals were evaluated analytically with our implementation described in Ref.~\citenum{baiardi_explicitly_2022}.

As a reference, FCI results for LiH were obtained with the CISDTQ implementation in the Psi4~\cite{smithSI4OpensourceSoftware2020} program with conventional integrals obtained from the restricted Hartree-Fock routine from Psi4.

\subsubsection{Effect of Normal-Ordered Approximation}

\begin{figure}[htb]
	\centering
	\includegraphics[width=0.9\linewidth]{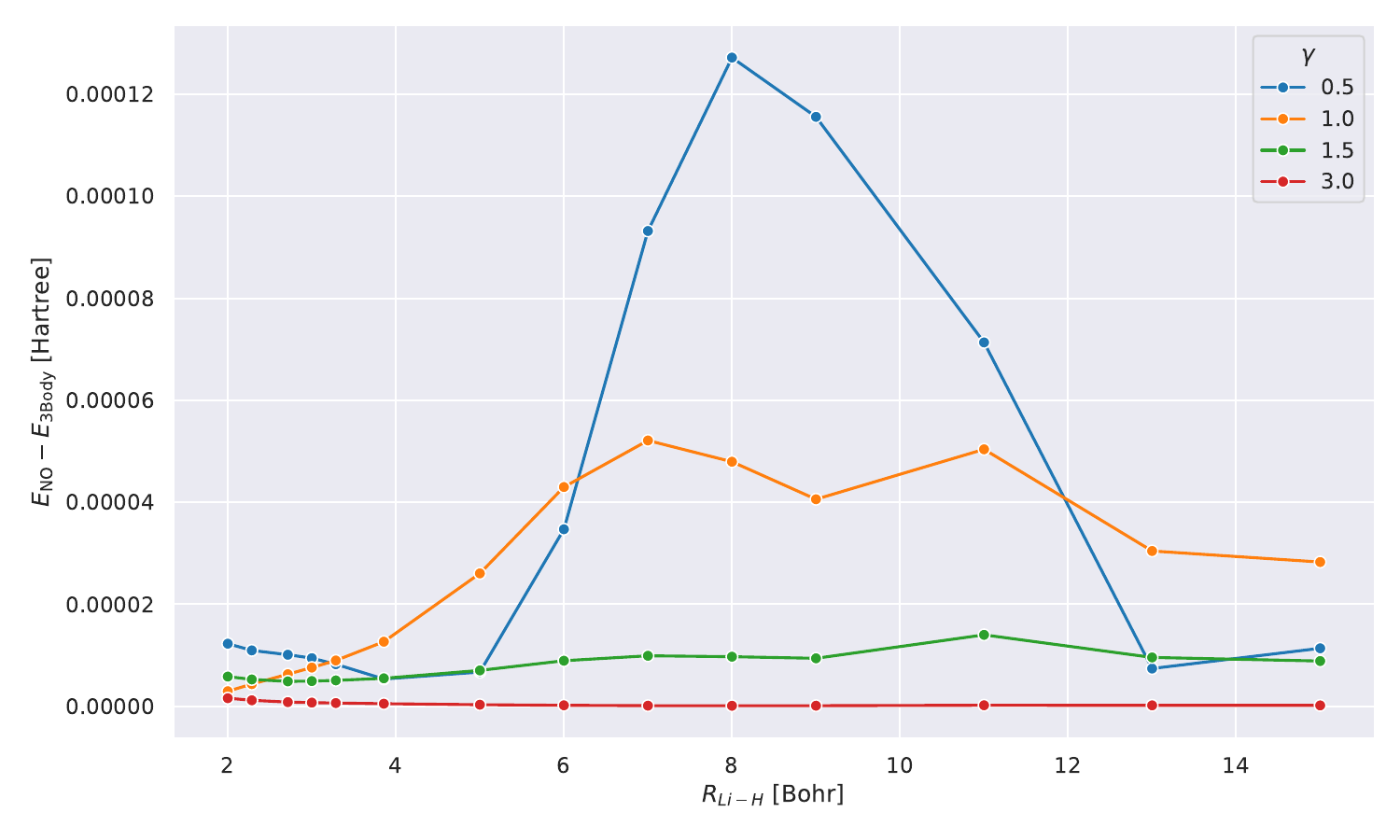}
	\caption{
    Error introduced by neglecting the pure three-body contribution in the normal-ordered $\hat{H}_{\mathrm{tc}}$ with respect to the Hartree-Fock determinant.
  The absolute error is given with respect to the electronic energy obtained with the full $\hat{H}_{\mathrm{tc}}$ along the potential energy curve of \ce{LiH}.
	Each curve corresponds to a tcDMRG calculation performed in the cc-pVDZ basis with a distinct value of $\gamma$ in the correlator $\tau = \frac{1}{2}\sum_{i < j} r_{ij}e^{-\gamma r_{ij}}$. }
	\label{fig:LiH-no-error}
\end{figure}

Fig.~\ref{fig:LiH-no-error} shows the absolute 
difference of the energies calculated with tcDMRG using the full treatment of the three-body operator and its normal-ordered approximation for different values of $\gamma$ along the potential energy curve.
We observe that the error resulting from this approximation decreases as the value of $\gamma$ increases.
This can be understood by recalling that as $\gamma \to \infty$, the effect of the transcorrelated similarity transformation vanishes.
Therefore, any approximation that is introduced in the treatment of $\hat{H}_{\mathrm{tc}}$ also disappears.
Our results suggest that this approximation is justified, as also shown in Ref.~\citenum{Christlmaier2023Jul} since the error is smaller than 0.13 mHartree for all values of $\gamma$.
Therefore, in all the following results, this approximation is applied.

\begin{figure}[htb]
    \centering
    \includegraphics[width=0.9\linewidth]{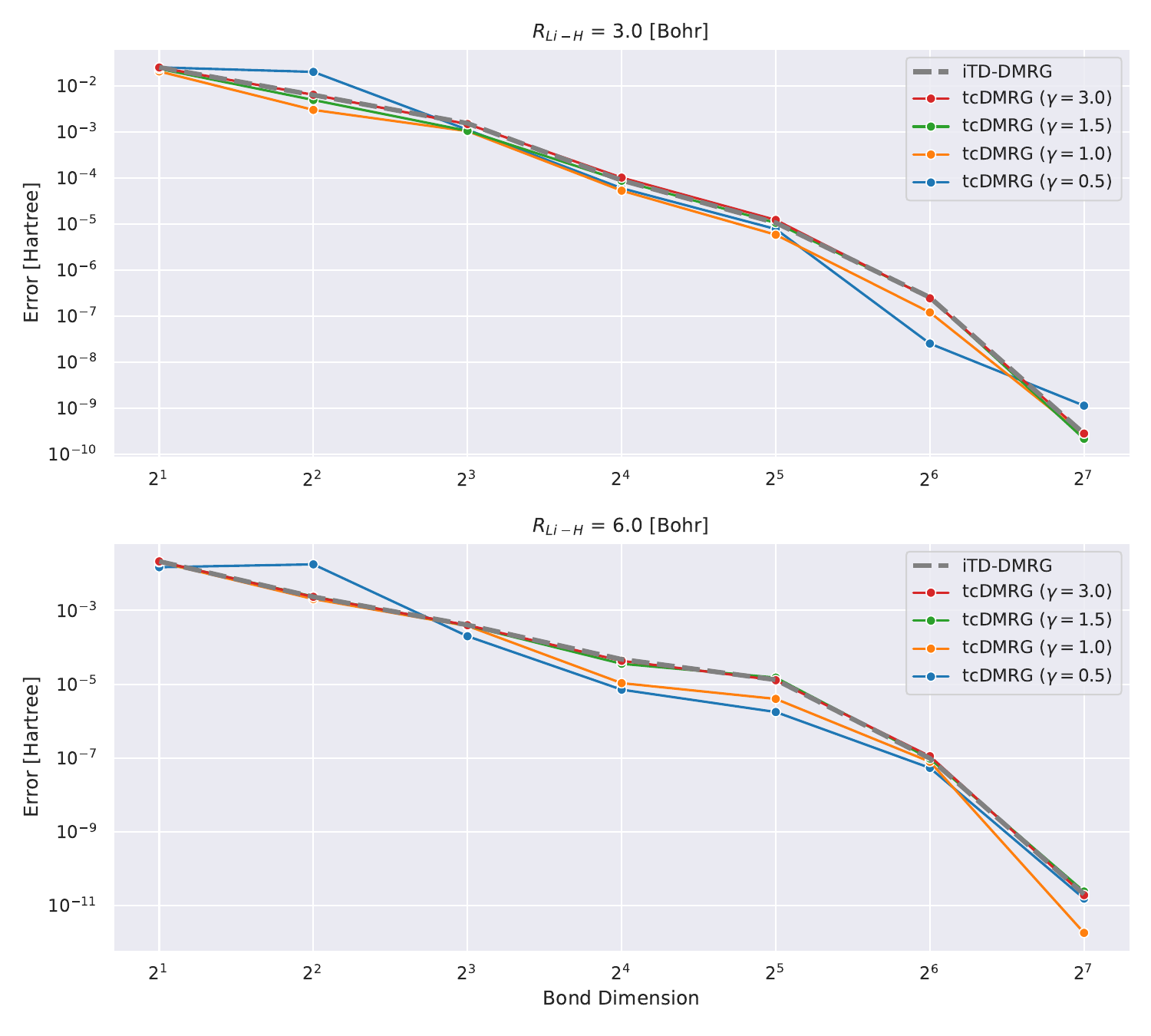}
    \caption{Convergence of the energy computed with DMRG for increasing bond dimension at two bond lengths, namely at the equilibrium of 3 Bohr (top) and a stretched geometry of 6 Bohr (bottom).
    All calculations were conducted using the cc-pVDZ basis, and the normal-ordered approximation was assumed for the tcDMRG computations.
    The error was measured with respect to a DMRG calculation with a bond dimension of $2^8=256$.
    }
    \label{fig:bond-dim}
\end{figure}

In Fig.~\ref{fig:bond-dim}, we analyze the convergence of the energy with increasing bond dimension in the DMRG calculation.
It appears that for this system, the tcDMRG method leads to either identical or smaller errors for bond dimensions larger than 4 than the conventional DMRG approach, depending on the parameter $\gamma$ in the correlator.
However, in general, the convergence behavior of the two methods remains similar.

\subsubsection{Effects of the Damping Parameter in the Correlator on Spectroscopic Quantities}

\begin{figure}[htb]
	\centering
	\includegraphics[width=0.9\linewidth]{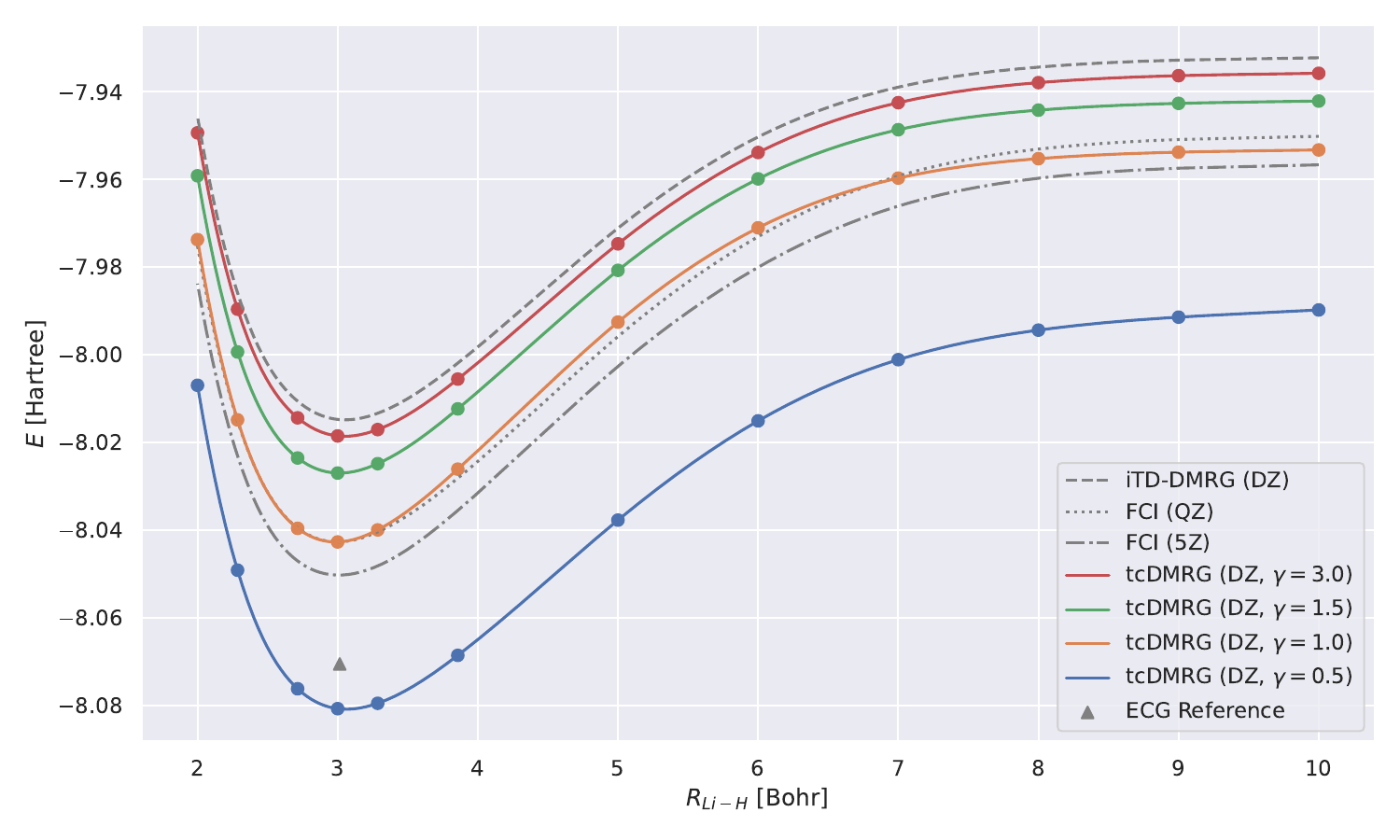}
	\caption{\ce{LiH} tcDMRG ground-state potential energy curves
		(cc-pVDZ (DZ) basis set, normal-ordered approximation) obtained with different values of $\gamma$.
		For comparison, imaginary-time time-dependent DMRG (iTD-DMRG) results
  in the cc-pVDZ basis, and FCI results obtained with the larger cc-pVQZ (QZ) and cc-pV5Z (5Z) basis sets are given.
    An explicitly correlated Gaussian (ECG) result~\cite{cencekBenchmarkCalculationsHe22000} is provided as a highly accurate estimate of the exact Born-Oppenheimer energy.
  }
	\label{fig:LiH-gammas}
\end{figure}

We now study the effect of the damping parameter $\gamma$ on the potential energy curves; the results are reported in Fig.~\ref{fig:LiH-gammas}:
Four tcDMRG curves were calculated with distinct values of $\gamma$ in the cc-pVDZ basis.
For comparison, the energies obtained with DMRG using the conventional electronic Hamiltonian in the cc-pVDZ basis.
This calculation was converged to the FCI solution and serves as a baseline on which the transcorrelated method should improve.
As a DMRG-independent reference, FCI solutions were calculated in the cc-pVQZ and cc-pV5Z basis sets~\cite{Dunning1989Jan}.
A single-point explicitly correlated Gaussian (ECG) energy, at a bond length of 3.015 Bohr, has been included from the literature~\cite{cencekBenchmarkCalculationsHe22000}, which is supposed to be within 10-20 $\mu$Hartree of the exact Born-Oppenheimer energy, making it a highly accurate estimate of the exact Born-Oppenheimer electronic energy.
When considering the FCI energies, even with the large quintuple-$\zeta$ basis set, the difference between them and the ECG result is of the order of 20 mHartree.
This highlights the accuracy of explicitly correlated methods.
However, in practice, relative energies and properties (e.g., energy derivatives) are often desired, for which determinantal expansions can benefit from error cancellation. Systematic error cancellation is key to the success of computational chemistry because the relative energies and properties are then affected by significantly smaller errors than what is observed here in absolute terms.

Turning our attention to the transcorrelated results for increasing values of $\gamma$, it is clearly seen that the effect of transcorrelation decreases in such a way that the electronic energies approach the ones obtained by conventional iTD-DMRG in the same one-particle basis (cc-pVDZ).
By contrast, we observe for small $\gamma$ a large effect on the electronic energies leading to solutions that even fall below the exact energy (represented by the ECG reference result of Ref.~\citenum{cencekBenchmarkCalculationsHe22000}), 
highlighted by the potential energy curve obtained for $\gamma=0.5$.
Reducing the value of $\gamma$ even further to 0.1, we found that the energy at an interatomic distance of 3 Bohr falls even to -8.578 Hartree, which is $\sim$0.5 Hartree below the exact energy.

For intermediate values of $\gamma$, especially for one which corresponds to an omission of this parameter, i.e., for  $\gamma=1.0$, we observe that the tcDMRG results coincide with the quadruple-$\zeta$ FCI results around the equilibrium distance, resulting in improved accuracy of two cardinal numbers (w.r.t. the zeta parameter of the orbital basis).
However, this improvement is not conserved along the entire potential energy curve, and ultimately, the tcDMRG curve undershoots the FCI results.

\begin{figure}[htb]
    \centering
    \includegraphics[width=0.9\linewidth]{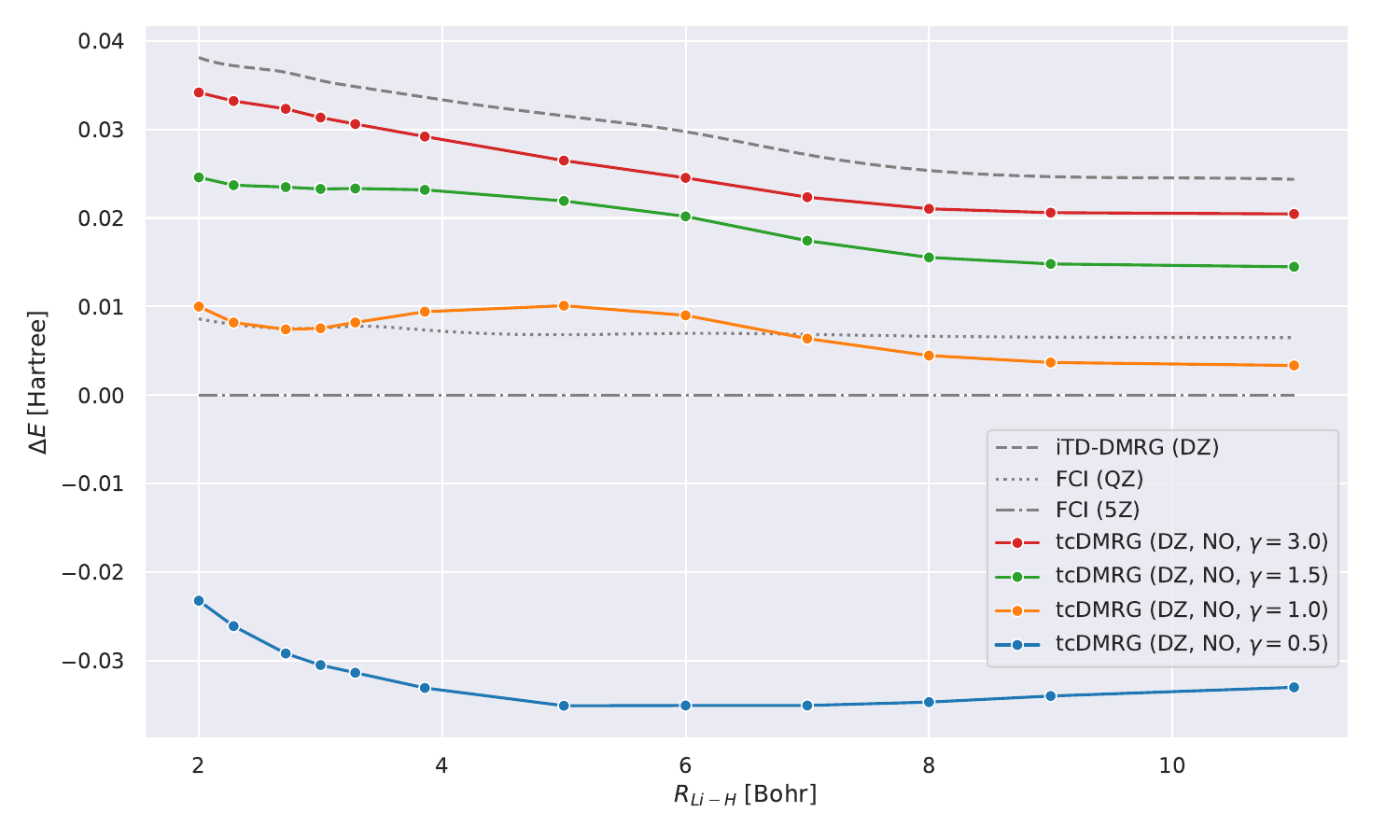}
    \caption{Error of ground state LiH potential energy curve with respect to the energies obtained using FCI calculations in the cc-pV5Z basis: $ \Delta E(R_{Li-H}) = E_{\text{method}}(R_{Li-H}) - E_{\text{FCI cc-pV5Z}}(R_{Li-H})$.
    The FCI quintuple-$\zeta$ curve is thus by definition a horizontal line at 0.
    }
    \label{fig:npe}
\end{figure}

We monitor the parallelity of the electronic energy obtained with various methods,
    \begin{equation}
    \Delta E(R_{Li-H}) = E_{\text{method}}(R_{Li-H}) - E_{\text{FCI cc-pV5Z}}(R_{Li-H}),
\end{equation}
measured against the quintuple-$\zeta$ FCI solution in Fig.~\ref{fig:npe}.
As can be seen, the energy error curves are not horizontal, indicating that the dissociation energy will be underestimated (see also Table~\ref{tab:Eeq-De}).
Moreover, the tcDMRG energy-error curves exhibit a hump which increases with increasing parameter $\gamma$ in the correlator.

Table~\ref{tab:Eeq-De} collects the total electronic energy at equilibrium distance, the electronic dissociation energies $D_e$, vibrational constants $\omega_e$ and $\omega_e x_e$ and the non-parallelity errors (NPE) obtained by the various methods.

\begin{table*}[htb!]
	\centering
	\begin{tabular}{lllllll}\toprule
		Method                                                  & $R_{eq}$ [Bohr] & $E(R_{eq})$ [Ha] & $D_e$ [eV] & $\omega_e$ [cm$^{-1}$] &  $\omega_e x_e$ [cm$^{-1}$] & NPE [Ha] \\
		\midrule
		Reference & 3.015~\cite{cencekBenchmarkCalculationsHe22000} & -8.070538~\cite{cencekBenchmarkCalculationsHe22000}    & 2.52~\cite{irikuraExperimentalVibrationalZeroPoint2007}  & 1405~\cite{irikuraExperimentalVibrationalZeroPoint2007} & 23~\cite{irikuraExperimentalVibrationalZeroPoint2007} & -            \\
		FCI (5Z)                                                & 3.005 & -8.050255 & 2.56 & 1369 & 12 & 0 (\textit{by def.})             \\
		FCI (QZ)                                                & 2.995 & -8.042676 & 2.53 & 1409 & 36 & 0.002093      \\
		FCI (TZ)                                                & 3.016 & -8.036660 & 2.47 & 1404 & 23 & 0.007405     \\
		FCI (DZ)                                                & 3.051 & -8.014803 & 2.26 & 1359 & 22 & 0.013682     \\
		iTD-DMRG (DZ)                                                & 3.051 & -8.014803 & 2.26 & 1359 & 22 & 0.013682     \\
		tcDMRG (DZ, $\gamma=3.0$)                               & 3.048 & -8.018552 & 2.25 & 1364 & 22 & 0.013566        \\
		tcDMRG (DZ, $\gamma=1.5$)                               & 3.009 & -8.026977 & 2.32 & 1408 & 25 & 0.009773      \\
		tcDMRG (DZ, $\gamma=1.25$)                               & 2.992 & -8.032831 & 2.37 & 1441 & 26 & 0.008475 \\
		tcDMRG (DZ, $\gamma=1.0$)                               & 2.983 & -8.042693 & 2.44 & 1476 & 26 & 0.006439      \\
		tcDMRG (DZ, $\gamma=0.75$)                               & 3.003 & -8.058122 & 2.53 & 1465 & 24 &  0.005913 \\
		tcDMRG (DZ, $\gamma=0.5$)                               & 3.061 & -8.080827 & 2.50 & 1354 & 22 & 0.011948     \\
		\bottomrule
	\end{tabular}
  \caption{Calculated properties for LiH.
  The reference explicitly correlated Gaussian equilibrium energy, evaluated at an interatomic distance of 3.015 Bohr, was taken from the literature~\cite{cencekBenchmarkCalculationsHe22000}. 
  All other reference values (first row) are experimental data~\cite{irikuraExperimentalVibrationalZeroPoint2007}.
  The harmonic vibrational frequencies $\omega_e$ and anharmonicities $\omega_e x_e$ 
  were obtained by fitting a fourth-order polynomial around the minimum of each potential energy curve and using the formulas from Ref.~\citenum{crawfordComparisonTwoApproaches1996}.
  For the estimation of the dissociation energy $D_e$, a bond length of 15 Bohr was taken as the dissociation limit.
  The NPE is, by definition, equal to 0 for the quintuple-$\zeta$ basis set.
  }
	\label{tab:Eeq-De}
\end{table*}

The NPE is defined as the difference between the maximum and the minimum error of the potential energy curves with respect to FCI/cc-pV5Z taken as reference,
\begin{equation}
    \mathrm{NPE} = \max_{R_{Li-H}} \Delta E - \min_{R_{Li-H}} \Delta E.
\end{equation}
As discussed in the previous paragraph, parallelity of the potential energy curves is a desired property so that the calculation of relative energies can benefit from error cancellation.

Table~\ref{tab:Eeq-De} illustrates that the tcDMRG method with large values of $\gamma$ converges to the conventional DMRG results in the same basis.
By decreasing the parameter $\gamma$ from 3.0 to 1.0, the downward trend of $\Delta E$ (also seen in Fig.~\ref{fig:npe}) is mitigated, but results in the introduction of a hump in the error which becomes more significant with decreasing $\gamma$.
For $\gamma=0.5$, a relative overestimation of the energy leads to an increase in the NPE.

Concerning the harmonic frequency $\omega_e$, we find all approaches to scatter around the experimental result by up to 10\%.
In fact, we see a strong dependence on the atomic-orbital basis set, where FCI harmonic frequency of the quintuple-$\zeta$ basis set significantly deviates from the convergence of FCI results obtained for the smaller basis sets. Inspecting the potential energy curves in these cases does not point toward any serious problem but demonstrates the sensitivity of the harmonic frequency to the local shape of the potential energy curve. 
Similar observations can be made for the FCI anharmonic constant $\omega_e x_e$ in the large basis sets, which deviate significantly from all other anharmonicities obtained, which are on the order of 10\% of the experimental value.

Regarding $D_e$ (see Fig.~\ref{fig:npe}), we see that increasing the parameter $\gamma$ leads to a decreased slope of $\Delta E$, but also introduces a hump.
Overall, the flatter profile results in an improved estimation of $D_e$ for larger values of $\gamma$.

\section{Conclusions and Current Challenges}

In this work, we reviewed the contribution of explicit electron correlation in the ansatz for the electronic wavefunction. Different approaches introduce inter-particle distances into correlation factors designed to alleviate the cusp problem of approximations to the electronic wavefunction based on single-particle basis functions (orbitals). 

Although electron-nucleus and electron-electron-nucleus cusps profoundly impact the electronic wavefunction and, hence, total electronic energy, we here argued that they may be neglected for an electronic structure model that yields reliable electronic energy differences due to changes in the valence region.
The resulting model features a rather simple correlator.
We examined the properties of this correlator in comparison with more elaborate correlators that also consider electron-nucleus distances explicitly (and hence, molecular structure in the Born-Oppenheimer approximation). We discussed that our simple electrons-only correlator allows us to define a molecular-structure-independent correlator and, therefore, leaves all structure dependence to be incorporated in the smooth function that is to be multiplied with the correlator. 

In our presentation, we highlighted the transcorrelated method's non-variationality and, hence, the importance of developing reliable variational methods applicable in this context.
We also addressed the source of the non-variationality of CC compared to the transcorrelated method.

The electronic correlator considers the analytic knowledge about the wavefunction near electron-electron cusps.
However, this behavior is to be suppressed at large inter-electronic distances, which is the reason for the introduction of an exponential that depends on this distance and acts as a damping function.
In principle, one may introduce a parameter that can be used to switch off the contributions of the inter-electronic distances so that one recovers results of the orbital basis without introduction of a correlator.
We argue that it should be possible to choose the parameter in this (nested) exponential in some simple, maybe even system-independent way.
This way, it will be guaranteed that a well-defined electronic structure model emerges that, for instance, does not rely on structure-dependent correlators that would change along a reaction coordinate or some other trajectory across a Born-Oppenheimer surface.

To highlight our arguments with a specific example, we provided results of the transcorrelated method (with the simple electrons-only correlator and fully analytic integral evaluation techniques) for the LiH diatomic molecule and related the results to accurate reference calculations.
We investigated the effect of the normal-ordering approximation of the three-body operator in the context of the tcDMRG method.
Moreover, we demonstrated that while the transcorrelated method has the potential to increase atomic orbital basis convergence, our current choice of correlator leads to non-systematic improvement in the energy along the potential energy curve.

In future work, a systematic analysis of the parameter in the damping function of the correlator should either reveal the best universal choice of this parameter valid for all molecules or provide analytic means to choose this parameter in a system-specific manner without the need for extensive prior optimization. Subsequently, it will be necessary to evaluate the still lacking long-range dynamic correlation that results from the choice of an active orbital space in a multi-configurational ansatz. Those correlations may then be efficiently captured by 
multi-reference-driven single-reference CC models~\cite{Oliphant1991,Piecuch_1993, kinoshitaCoupledclusterMethodTailored2005,Veis2016Oct,Morchen2020Dec,Faulstich_2019},
which might provide higher accuracy results than multi-reference perturbation theory would deliver for a transcorrelated zeroth-order Hamiltonian.

\section*{Author Contributions}
\textbf{Kalman Szenes}: Conceptualization, Methodology, Investigation, Software, Writing - original draft, Validation.
\textbf{Maximilian M\"orchen}: Conceptualization, Methodology, Investigation, Software, Writing - original draft, Validation.
\textbf{Paul Fischill}: Software.
\textbf{Markus Reiher}: Conceptualization, Methodology, Validation, Resource, Writing - original draft, Writing - review and editing, Supervision, Project administration, Funding acquisition.

\section*{Conflicts of interest}
There are no conflicts to declare.

\section*{Acknowledgments}
This work was supported by the Swiss National Science Foundation (No. 200021\_219616) and by ETH Research Grant ETH-43 20-2.



\balance


\providecommand*{\mcitethebibliography}{\thebibliography}
\csname @ifundefined\endcsname{endmcitethebibliography}
{\let\endmcitethebibliography\endthebibliography}{}

\end{document}